\documentstyle[preprint,aps]{revtex}
\tightenlines
\input psfig.sty
\begin{document}
\title { Violations of Lorentz Covariance in Light Front Quark Models}
\author{K. Bodoor$^1$, H. J. Weber$^1$, T. Frederico$^2$, and M. Beyer$^3$}
\address{$^1$Institute of Nuclear and Particle Physics, University of
Virginia, \\Charlottesville, VA 22904-4714, USA\\ $^2$Dep. de F\'isica, 
Instituto Tec. de Aeron\'autica, Centro T\'ec. Aerospacial\\ 12.228-900 S\~ao 
Paulo, Brazil\\ $^3$ Fachbereich Physik, Universit\"at Rostock, 18051 Rostock, 
Germany}
\maketitle
\begin{abstract}
Electromagnetic form factors of the nucleon from relativistic quark 
models are analyzed: results from null-plane projection of the Feynman 
triangle diagram are compared with a Bakamjian-Thomas model. The magnetic 
form factors of the models differ by about 15\% at spacelike momentum transfer 
$0.5~\mathrm{GeV}^2$, 
while the charge form factors are much closer. Spurious contributions to 
electromagnetic form factors due to violations of rotational symmetry are 
eliminated for both models. One method changes magnetic form factors by 
about $10\%$, whereas the charge form factors stay nearly the same. Another 
one changes the charge form factor of the Bakamjian-Thomas model by more than 
$50\%$.         
\end{abstract}
 \par   
\newpage

\section{Introduction }
In light-cone quark models wave functions may be boosted kinematically, that 
is, independent of interactions. Because form factors involve boosted wave 
functions, this attractive feature has motivated many electroweak form factor  
calculations~\cite{BTel,NPel} in Dirac's light-front form of relativistic 
few-body physics~\cite{PAMD}. When the quark model is Lorentz invariant and 
the exact (interaction dependent) current is evaluated, the form 
factors will be Lorentz invariant. Because most form factor models are based 
on free quark currents (in impulse approximation, IA, and/or other 
approximations), the resulting form factors are in fact frame dependent.    
\par
The light-front form is obtained from the instant form in the 
infinite momentum limit, and this amounts to the well-known change of momentum 
variables $(p^+=p^{0}+p^z,p^-=p^{0}-p^z)$ on the light cone~\cite{SK}. Here 
the conventional choice of the light cone direction is the z-axis and the 
light cone time is $x^+=t+z$. Because in front form both transverse rotation 
generators become interaction dependent, rotational symmetry is more difficult 
to implement. Thus light front models include kinematic boosts at the 
expense of interaction dependent angular momentum operators. 
\par
Light front models obtained from quantum field theories usually involve 
interacting spin operators. There is a unitary (but interaction 
dependent) transformation to a representation where spins are free and state 
vectors are specified on a null plane $\omega \cdot x=0$ with a lightlike 
(i.e. $\omega^2=0$) four-vector $\omega^\mu$~\cite{KF}. If a theory is Lorentz 
invariant and the exact current operator is known, e.g. in quantum field 
theory given by the Mandelstam prescription, its form factors are 
expected to be independent of $\omega^\mu$. In this case the electromagnetic 
current operator in general has many-body contributions~\cite{FL}. When it is 
approximated by a sum of free one-body currents, as in the impulse 
approximation for example, Lorentz invariance of the form factors is violated. 
 
When Lorentz invariance of the form factors is violated, additional 
$\omega$-dependent current components arise along with their form factors 
which are needed to fully parameterize the current matrix elements that now 
obey fewer constraints. When Lorentz invariance of the quark model form 
factors is fully restored by including two-body currents, these spurious 
currents are canceled and the remaining calculated form factors are modified 
as well. In other words, achieving $\omega$-independence of form factors by 
eliminating spurious from factors is just one step towards Lorentz invariance 
but, by itself, it does not guarantee correct form factors yet. In the case of 
$\omega$-dependence the unphysical currents can be removed. As a bonus, the  
inconsistencies between current matrix elements of different helicities are  
reduced. This is done here for electromagnetic nucleon form factors of 
constituent quark models in light front dynamics. These models are kept as 
simple as possible (e.g. no anomalous magnetic moments of constituent quarks) 
because we are not attempting to fit the data. Rather we wish to compare their 
electromagnetic form factors and the size of errors due to violations of the  
rotational symmetry.     

Two light-cone versions of the constituent quark model (NQM), one based on the 
Bakamjian-Thomas prescription~\cite{BT}, called BT in the following, and 
another on the null-plane projection of a triangle Feynman diagram (see 
Ara\'ujo {\it et al.} in~\cite{NPel}), called KW (see Konen {\it et al.} 
in~\cite{NPel} for $\alpha=1/2$), are briefly reviewed in Sect.II. The 
electromagnetic current matrix elements are given in Sect.III, and the 
results in Sect.IV and conclusions are presented in Sect.V. 

\section{Light-Cone Quark Models}

We start with a brief description of the light front notation. 
The longitudinal quark momentum fractions are defined as $x_i=p^+_i/P^+$ with 
the total nucleon momentum $P^+=\sum_{i}p^+_i$ so that $\sum_{i}x_i=1$ and 
$0\leq x_i\leq 1$, where $p_i$ are the quark momentum variables. The $i$th 
quark momentum in the nucleon rest frame is given by 
\begin{equation}
\vec {k}_{i\perp}=\vec {p}_{i\perp} -x_i \vec {P}_\perp,  
\end{equation}
with $\vec k_{\perp}=(k_x,k_y)$ etc. 
so that these internal variables satisfy $\sum_i \vec{k}_{i\perp}=0$ and 
$k^+_i=0$. For a three-quark bound state the relative four-momentum variables 
are the space-like Jacobi momentum variables in which 
the kinematic invariants $x_i$ play the role of masses. For example, $q_3$ is 
the relative quark momentum between the up quarks of the proton in the 
uds-basis and $Q_3$ between the down quark and the up 
quark pair, so that for the $+$ and $\perp=(x,y)$ components 
\begin{eqnarray}
q_3={x_1p_2-x_2p_1\over x_1+x_2},\quad Q_3=(x_1+x_2)p_3-x_3(p_1+p_2), 
\label{jac}
\end{eqnarray}
etc. Because $q_3^+=0=Q_3^+$ both relative momentum variables are space-like. 
In the uds-basis~\cite{JF} quarks are treated as distinguishable, and up 
and down quarks are symmetrized explicitly. The antisymmetric color wave 
function is suppressed. For the proton (neutron) the down (up) quark carries 
the label $3$.  

In light front dynamics the total momentum motion rigorously separates 
from the internal motion. Therefore, the internal nucleon wave function 
$\psi(x_i, q_3, Q_3, \lambda_i) $ does not change under kinematic Lorentz 
transformations or translations. Here the $\lambda_i$ are the quark helicities. 
Thus, if the wave function is known in the nucleon rest frame it is known 
everywhere because the seven kinematic generators are transitive in the null 
plane.   
\par
Three-quark wave functions for baryons have been constructed in light cone 
versions of the constituent quark model (LCQM) to study the static properties 
of the nucleon~\cite{BTel,NPel} and electromagnetic transition form 
factors for $N \to N^*$ and $N \to\Delta$ processes~\cite{BTin,NPin}. 
Such relativistic quark models significantly improve many predictions of the 
NQM of which the nucleon weak axial charge $g_A$ is the best known example. 
For a recent analysis of the static observables and nucleon electroweak form 
factors see ref.~\cite{FBW}. 

The three-quark wave function for a nucleon is the product of a
totally symmetric momentum wave function and a nonstatic spin wave function 
$\chi_\lambda$ which is an eigenfunction of the total angular momentum 
(squared) and its projection on the light cone axis. The spin wave function 
can be represented as a linear combination of products of matrix elements 
between valence quark light-cone spinors coupled by appropriate $\gamma$ (and 
isospin)-matrices to the spin and isospin of the nucleon. 
Free Melosh rotations~\cite{M} are of central importance in LCQMs for the 
construction of relativistic many-body spin-flavor wave functions for 
hadrons from those of the nonrelativistic quark model (NQM). Instant 
Dirac-spinors are transformed into light-front helicity eigenstates 
(light-cone Dirac-spinors) by Melosh transformations which include 
the kinematic quark boosts. Direct coupling of Pauli spinors by SU(2) 
Clebsch-Gordan coefficients leads to a set of spin-flavor wave functions 
that are in one-to-one correspondence with the NQM states. (For details, see 
ref.~\cite{BKW}.)    
\par
The wave function of the nucleon with nucleon helicity $\lambda$ is here taken 
to be 
\begin{equation}
\psi_N(\lambda)=N\phi(x_3,q_3,Q_3)\chi_\lambda, 
\label{nwf}
\end{equation}
where $N$ is a normalization constant fixed by the proton charge (electric form 
factor at $q^2=0$). The Gaussian momentum wave function is written in terms 
of the relative momentum variables according to the Brodsky-Huang-Lepage 
prescription~\cite{BHL} 
\begin{equation}
\phi(x_i,q_3,Q_3)=e^{-M_3^2/6\beta^2}
\label{gwf} 
\end{equation}
with a size parameter $\beta$ and the free mass operator $M_3$, i.e. the sum 
of the quark light-cone energies in the nucleon rest frame,   
\begin{eqnarray}
M_3^2=\sum_{i=1}^3 {\vec {k}_{i\perp}^2 +m_i^2  \over x_i}
     =-q_3^2{1-x_3 \over x_1x_2}-{Q_3^2 \over x_3(1-x_3)}+
\sum_{i=1}^3 {m_i^2  \over x_i}.
\label{m3}
\end{eqnarray}

In the projection of the triangle Feynman diagram to the null plane the 
totally symmetric nonstatic proton spin wave function $\chi_\lambda$ 
(see~\cite{BKW} for more details) originates from the quark-nucleon 
Lagrangian~\cite{FBW}   
\begin{eqnarray}
{\cal L}_{q^3-N}=\left(\bar q_1 [i\partial_N]\gamma_5 i\tau_2C\bar 
q^{T}_2\right)\cdot \bar q_3\Psi_N + (23)1+(31)2,
\label{qnc}
\end{eqnarray} 
where $q_i$ and $\Psi_N$ are the quark and nucleon fields, 
$C=i\gamma^2\gamma^0$ is the charge-conjugation operator, and $\gamma_5$ 
is characteristic of the spin $0$ of the quark pair~\cite{BD}. The term 
$[P]\equiv {\gamma \cdot P +m_N \over 2m_N}$ with $P$ the total momentum and 
$m_N$ the nucleon mass for 
the null-plane projection model (KW), while for the Bakamjian-Thomas model 
(BT) the nucleon mass in $[P]$ is replaced by the free three-quark mass $M_3$ 
of Eq.~\ref{m3}. In the BT-model $[P]$ derives from the Melosh rotation of the 
quark Pauli spinors. The nucleon positive-energy projection operator $[P]$ 
corresponds to the mixing parameter $\alpha=1/2$ in Ara\'ujo 
{\it et al.} of~\cite{NPel} of two of the three linearly independent 
three-quark-nucleon couplings that reduce to the NQM S-wave ground state in 
the static limit~\cite{BKW,HJW}. We restrict ourselves to the case 
$\alpha=1/2$ because it can be directly compared with the BT-model.  

In the triangle diagram for the form factors the quark fields are replaced by 
light-cone spinors $u_i$, quark 3 being the down quark of the proton in the 
$uds$-basis, so that the conventional nonstatic spin 
wave function 
\begin{eqnarray} 
\chi_N=\bar {u}_2 [P] \gamma_5 C \bar {u}_3^T \cdot \bar {u}_1 u_N - 
\bar {u}_3 [P] \gamma_5 C \bar {u}_1^T \cdot \bar {u}_2 u_N  
\label{swf}
\end{eqnarray}
results from Eq.~\ref{qnc}. The momentum wave function $(M_3^2-m_N^2)\phi$ in 
Eqs.~\ref{nwf},\ref{gwf} is proportional to the vertex function of 
${\cal L}_{q^3-N}$ of Eq.~\ref{qnc}.     

These simple relativistic quark models depend on two parameters, the 
common $u, d$ constituent quark mass $m_q$ and the size constant 
$\beta $ of the confinement potential. The proton mean square radius 
determines $\beta $ so that $1/\beta \sim \langle r^2\rangle_P^{1/2}$ up to 
relativistic corrections. 

\section{Electromagnetic Form Factors of the Nucleon} 

If $J^{\mu}$ is the full electromagnetic current, the current matrix 
element of a nucleon state consists of the Dirac and Pauli conserved 
currents,
\begin{equation}
\langle N(P',\lambda')|J^{\mu}|N(P,\lambda)\rangle
=\bar {u}_N(P',\lambda') [\gamma^{\mu} F_1(q^2) + 
{i \sigma^{\mu \nu} q_{\nu} \over 2m_N} F_2 (q^2) ] 
u_N(P,\lambda),       
\label{ncur}
\end{equation}
where $\sigma^{\mu \nu}=i[\gamma^{\mu},\gamma^{\nu}]/2$. 
We will also use the standard electric and magnetic Sachs form factors 
\begin{eqnarray}
G_E(q^2)=F_1(q^2)-\eta F_2(q^2),\quad G_M=F_1+F_2,  
\end{eqnarray}   
where $\eta\equiv -q^2/4m^2_N\geq 0$.
\par
If free quark currents are used as in the impulse approximation, the possible 
$\omega$-dependence of the 
electromagnetic current matrix elements between baryon states leads to three 
additional conserved components~\cite{KM} with form factors $B_i$ so that 
Eq.~\ref{ncur} becomes 
\begin{eqnarray}
\langle N(P',\lambda')|\sum_j e_j\bar q_j \gamma^{\mu}q_j |N(P,\lambda)\rangle
&=&\bar {u}_N(P',\lambda')\left( F_1 \gamma^{\mu}+i{F_2\over 2m_N}\sigma^{\mu 
\nu}q_\nu \right)u_N(P,\lambda)\label{iac}\\
&&+\bar {u}_N(P',\lambda')B_1\left({\omega\cdot \gamma \over \omega\cdot P}
-{1\over (1+\eta)m_N}\right)(P'^{\mu}+P^{\mu})u_N(P,\lambda)\nonumber\\
&&+\bar {u}_N(P',\lambda')\left(B_2{m_N\over \omega\cdot P}\omega^{\mu}
+B_3{m_N^2\over (\omega\cdot P)^2}\omega \cdot \gamma \omega^{\mu}\right)
u_N(P,\lambda),\nonumber 
\end{eqnarray}
where $\omega\cdot q=q^+=0$ for the standard choice of light cone axis.    

In front form all form factors are usually extracted from matrix 
elements of $J^+$ in the light-cone analog of the Breit frame (where $q^+=0$), 
the so-called 'good' current component~\cite{LS} 
\begin{eqnarray}\nonumber
eF_1(q^2)&=&{m_N\over P^+}\langle N(P')_{\uparrow}|J^+|N(P)_{\uparrow}\rangle,
\\{eq^L\over 2m_N}F_2(q^2)&=&-{m_N\over P^+}\langle N(P')_{\uparrow}|J^+|
N(P)_{\downarrow}\rangle, 
\label{ff}
\end{eqnarray}
with $q^L=q_x-iq_y$.
This is no longer possible when $\omega$-dependent currents are 
present~\cite{XW}. We therefore define the more general current matrix 
elements 
\begin{equation}
J^{\mu}_{\lambda' \lambda}\equiv \langle N',\lambda'|J^{\mu}|N,\lambda\rangle 
\end{equation} 
in the rest frame of the nucleon. Just as Eq.~\ref{ff} follows from 
Eq.~\ref{ncur} we obtain 
\begin{eqnarray}\nonumber
J^+_{\uparrow \uparrow}&=&P^+\left({F_1\over m_N}+2B_1 \left[{1\over m_N}
-a\right]\right),\\
J^+_{\uparrow \downarrow}&=&P^+q^L\left(-{F_2\over 2m^2_N}-{aB_1\over m_N}
\right),\nonumber\\
J^L_{\uparrow \uparrow}&=&q^L\left({F_1\over m_N}+{F_2\over 2m_N}+B_1
\left[{1\over m_N}-a\right]\right),\nonumber\\
J^L_{\uparrow \downarrow}&=&(q^L)^2\left(-{F_2\over 4m^2_N}-{B_1a\over 2m_N}
\right),\nonumber\\
J^L_{\downarrow \uparrow}&=&-q^2\left({F_2\over 4m^2_N}+{B_1a\over 2m_N}
\right),\nonumber\\
J^L_{\downarrow \downarrow}&=&q^L\left(-{F_2\over 2m_N}+B_1\left[{1\over m_N}
-a\right]
\right)
\label{cme}
\end{eqnarray}
from Eq.~\ref{iac} and \ref{cme}, where $a=1/(1+\eta)m_N$. It is 
straightforward to check the symmetries 
\begin{eqnarray}
J^+_{\downarrow \downarrow}=J^+_{\uparrow \uparrow},\qquad 
{J^+_{\downarrow \uparrow}\over q^R}=-{J^+_{\uparrow \downarrow}\over q^L},
\end{eqnarray}
etc. From Eq.~\ref{iac} in conjunction with Eq.~\ref{cme} we find two 
solutions for the Dirac and Pauli form factors 
\begin{eqnarray}
F_1&=&m_N\{(2a m_N-1){J^+_{\uparrow \uparrow}\over P^+}+2m^2_N\left[{1\over 
m_N}-a\right]{J^+_{\uparrow \downarrow}\over P^+q^L}+2(1-a m_N){J^L_{\uparrow 
\uparrow}\over q^L}\},\nonumber\\
F_2&=&-m^3_N\{{2a\over m_N}{J^+_{\uparrow \uparrow}\over P^+}+2\left[{1\over 
m_N}-a\right]{J^+_{\uparrow \downarrow}\over P^+q^L} -{2a\over m_N}
{J^L_{\uparrow \uparrow}\over q^L}\},
\label{set1}
\end{eqnarray}
which we label set 1, while 
\begin{eqnarray}
F_1&=&m_N\{{J^+_{\uparrow \uparrow}\over P^+}+4m^2_N\left[{1\over m_N}-a\right]
\left({-J^L_{\downarrow \downarrow}\over 2m_Nq^L}+{J^L_{\uparrow \downarrow}
\over (q^L)^2}\right)\},\nonumber\\
F_2&=&-4m^2_N \{ [1-a m_N]{J^L_{\uparrow \downarrow}\over (q^L)^2}+{a\over 2}
{J^L_{\downarrow \downarrow}\over q^L}\}. 
\label{set2}
\end{eqnarray}
is the set 2. 

Subtracting the spurious form factors according to the prescriptions of 
Eqs.~\ref{set1},\ref{set2} will be called corrected form factors in the 
following.   

\section{Numerical Results}

The form factor calculations involve matrix elements of products of up to 
ten Dirac spinors which are evaluated and summed using symbolic codes 
generated by Mathematica. For the    
KW-model the transverse momentum integrations are carried out analytically, 
so that only the integrations over the longitudinal momentum fractions 
$x_1, x_2$ are evaluated numerically. The BT-model involves six-dimensional 
numerical integrations which are performed by Monte Carlo routines~\cite{PL}.  

Although relativistic models yield predictions for any value of the 
momentum transfer, the constituent quark concept is expected to lose validity 
(i.e. the momentum dependence of the quark selfenergy becomes less important)  
at the spontaneous chiral symmetry breaking scale ($\chi=4\pi f_\pi\sim 1 
\mathrm{GeV}$), where current quarks are expected to become the relevant 
degrees of freedom. This limits the validity of the models to fairly small 
values of $Q^2=-q^2$. Both of our light-cone quark models are based on 
pointlike constituent quarks. (Better fits can be obtained using small 
anomalous magnetic moments and a finite size form factor, but this is not our 
objective here. See Cardarelli {\it et al.} in~\cite{BTel}.)

In figures 1 to 4 we compare the electromagnetic form factors of the KW and 
BT models (uncorrected and corrected) for the {\bf same} parameters 
$\beta=0.32~\mathrm{GeV}$ and $m_q=0.32~ \mathrm{GeV}$. Results are shown 
relative to the dipole form factor $F_D=(1-q^2/0.71~GeV^2)^{-2}$ with $q^2$ in 
$\mathrm{GeV}^2$ except for Fig.~3 (neutron charge form factor). 
While the {\em charge} form factors of KW and BT (both uncorrected) for the 
proton in Fig. 1 are close to each other below $0.7~\mathrm{GeV}^2$, both 
models' {\em magnetic} form factors in Fig.~2 differ by a roughly 
constant amount $0.2$, i.e. $20\%$ at $1~\mathrm{GeV}^2$. The corrected 
magnetic proton and neutron form factors BT1 and BT2 in Figs.~2 and 4 are 
close to each other but differ by about $20\%$ from the uncorrected BT-model. 
This we interpret to mean that the corrected magnetic form factor of the 
BT-model is nearly unambiguous and correct. The corrected KW1 and KW2 form 
factors differ by a few percent (Figs.~2 and 4), however the correction 
itself is smaller than in the BT case. The charge form factors KW1 and KW 
coincide, so that the dotted line is not visible in the figures and KW2 leads 
to a slight correction in the proton case and a larger correction for the 
neutron. For the proton {\em charge} form factor BT1 and BT2 differ much more, 
so that it is poorly predicted by the BT-model in contrast to the KW-model.    
 
Better fits can be obtained for both models, but 
for different parameters sets, and samples are shown in Figs.~5-9. In contrast 
to the NQM, the proton charge form factor $G^p_E$ falls faster than $G^p_M$ in 
both relativistic quark models. This is a general feature of LCQMs, which is 
in qualitative agreement with the recent data from Jefferson 
Laboratory~\cite{Jlab}. Yet the proton charge form factors of both models fall 
off too fast compared to the JLab data in Fig.5.    

The major disagreement involves the neutron charge 
form factor that is low by a factor of more than $2$. However, this is not 
surprising because it is known to be sensitive to the spin force and 
pion cloud effects that are beyond the scope of the simple relativistic quark 
models considered here.         
    
The trends of uncorrected form factors persist when spurious form factors are 
subtracted according to Eqs.~\ref{set1},\ref{set2} as shown in Figs. 1 to 9. 
From Fig.~1 we see that the proton charge form factors of the KW-model and the 
BT model for set 1 are hardly affected by the spurious contributions, while 
the corrections for the BT-model for set 2 are more than $50\%$. For the 
magnetic nucleon form factors in Figs.~2, 4 the corrections are similar for 
both models.   
  
\section{Summary and Conclusion}

The impulse approximation leads to violations of gauge invariance and 
rotational symmetry by LCQMs, i.e. spurious dependence on the choice of 
light-cone axis. Depending on which helicity matrix elements are used to 
extract electromagnetic form factors, there are different schemes of 
subtracting spurious components from these form factors. 
Subtracting spuriosities reduces inconsistencies between helicity matrix 
elements. When electromagnetic form factors roughly agree that are obtained 
from different correction schemes, then we consider them as essentially 
correct and reliably predicted by the LCQM in question. This is the case 
for the magnetic proton and neutron form factors of the KW- and BT-models. 
The KW-model also predicts $G^p_E$ reliably in this sense, in contrast to 
the BT-model that predicts $G^p_E$ poorly. The typical corrections for 
spuriosities are larger for the BT-model as a consequence of its larger 
high-momentum components that enter via the three-quark mass $M_3$ in the 
quark-nucleon coupling in Eq.~\ref{qnc}. (Note that $M_3$ of Eq.~\ref{m3} 
depends on the internal quark-momentum variables.)  
For $-q^2<1\ \mathrm{GeV}^2$, the corrected form factors differ from the 
uncorrected ones by about $20\%$ for the KW-model and up to $50\%$ for the 
BT-model. These are also the typical differences between reliably predicted 
form factors of these models. Thus, the theoretical errors of the form factors 
of the models are of roughly the same size as the differences between 
reliably predicted form factors of the models, so that it is difficult to 
distinguish them. These findings suggest that it is important for LCQMs 
to improve the impulse approximation and use conserved electromagnetic 
currents that are consistent with the interactions of the models.    

\section{Acknowledgement}
KB gratefully acknowledges the support of UVa's INPP, TF thanks CNPq and FAPESP 
and MB also FAPESP for support. 


\tightenlines

\begin{figure}[h]
\vglue 2in
\centerline{\psfig{figure=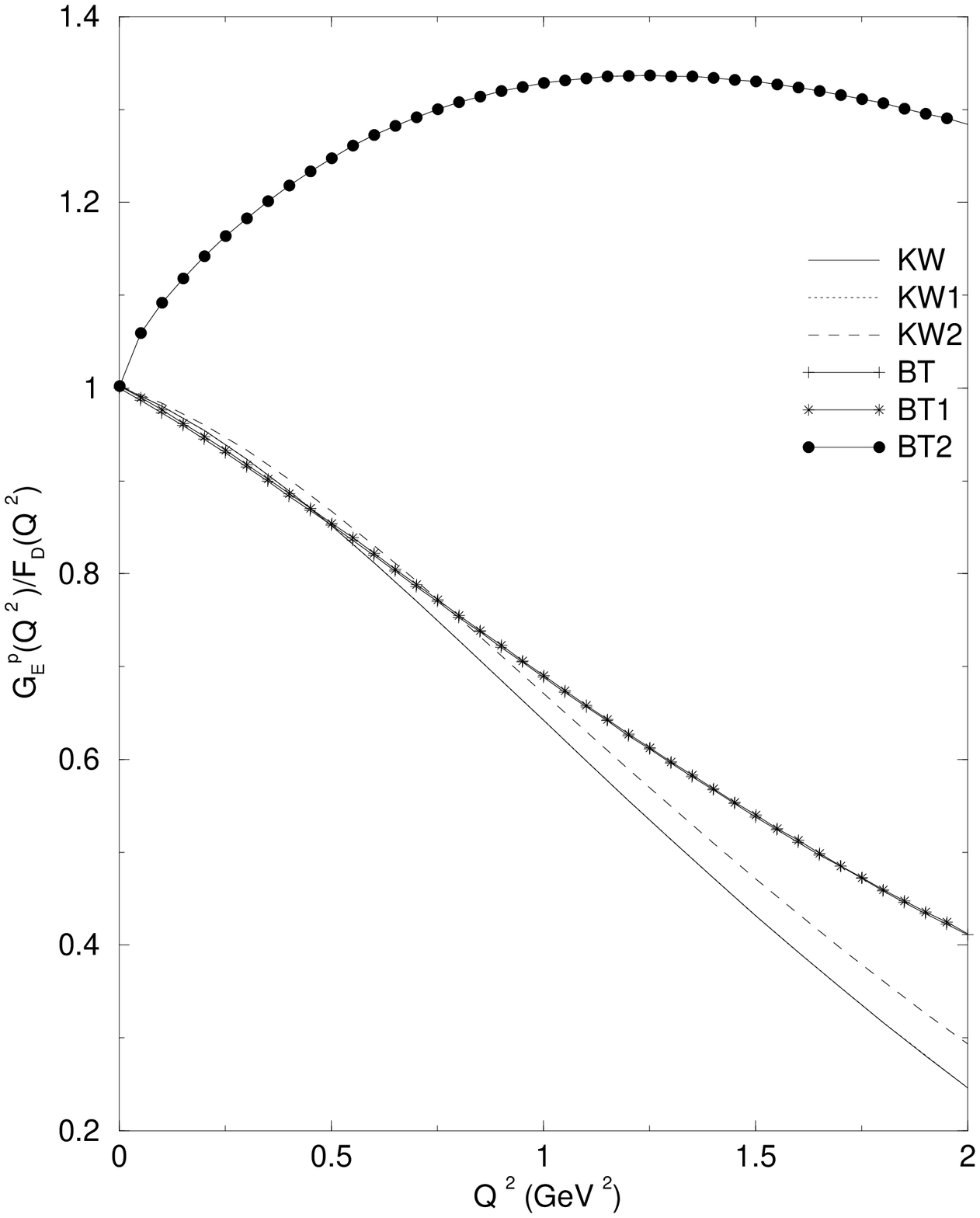,width=4in}}
\vglue 2in
\caption{Proton charge form factor $G^p_E/F_D$ relative to 
dipole shape $F_D$ for $m_q=\beta=0.32$\ GeV: The solid line is KW 
uncorrected while dashed and dotted lines KW1, KW2 are the corrected set 1,2 
cases, respectively. The BT-model curves are similarly denoted by BT, BT1, BT2.}
\end{figure}

\begin{figure}[h]
\vglue 2in
\centerline{\psfig{figure=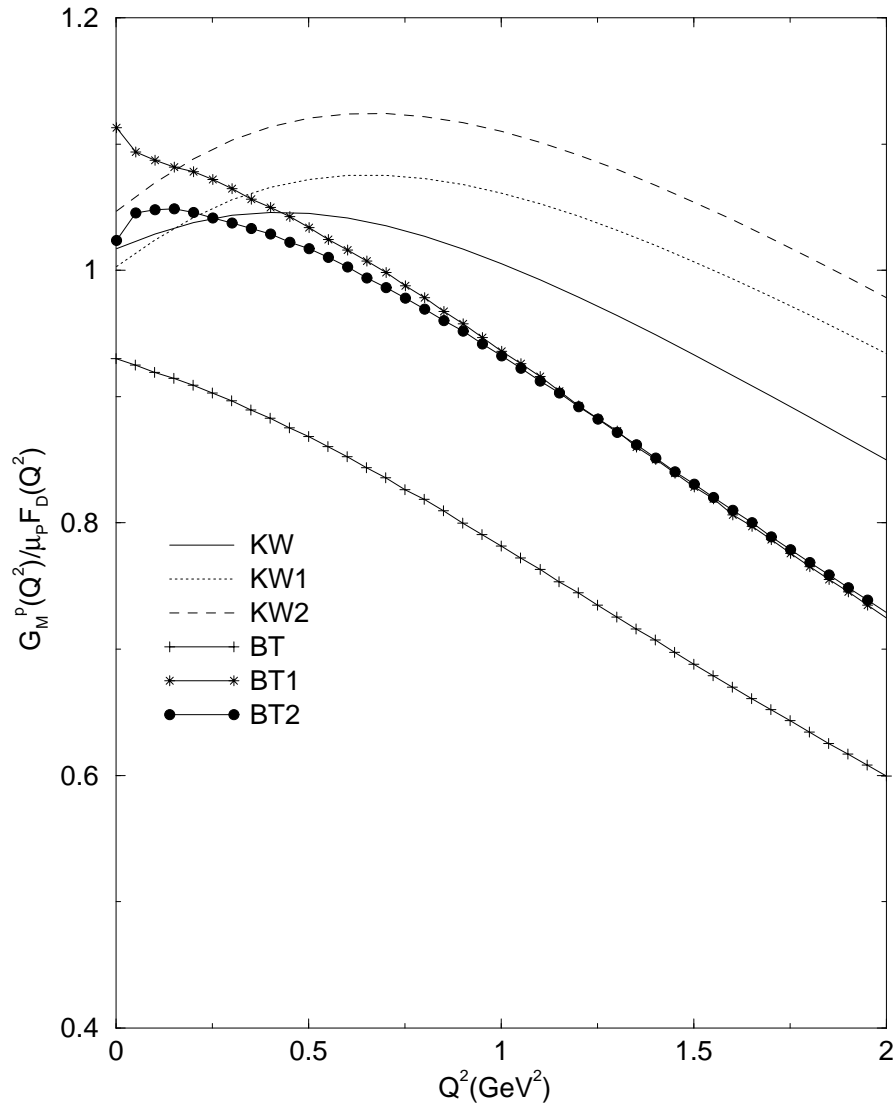,width=4in}}
\vglue 2in
\caption{Proton magnetic form factor $G^p_M/\mu_p F_D$ relative to 
dipole shape $\mu_p F_D$. The notation and parameters are as in Fig.~1.} 
\end{figure}

\begin{figure}[h]
\vglue 2in
\centerline{\psfig{figure=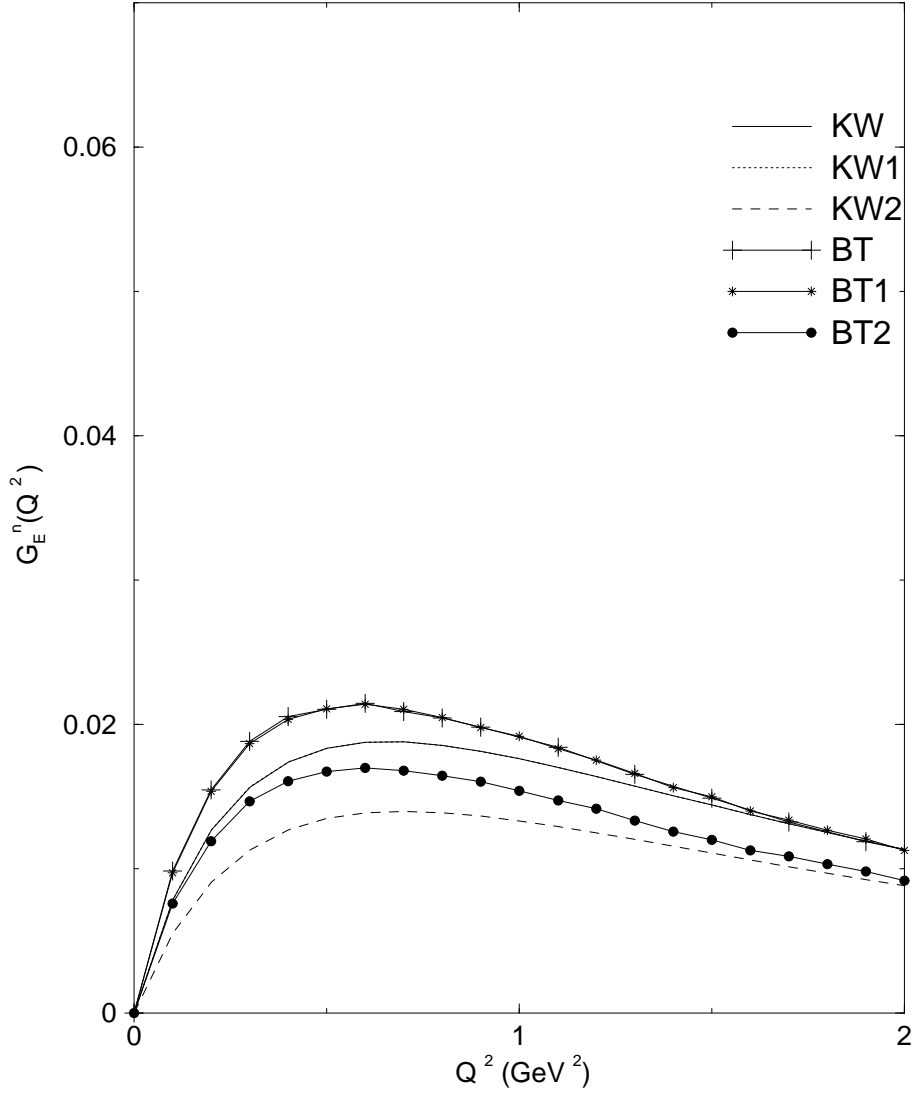,width=4in}}
\vglue 2in
\caption{Neutron charge form factor $G^n_E$ with the notation and parameters 
of Fig.~1. } 
\end{figure}

\begin{figure}[h]
\vglue 2in
\centerline{\psfig{figure=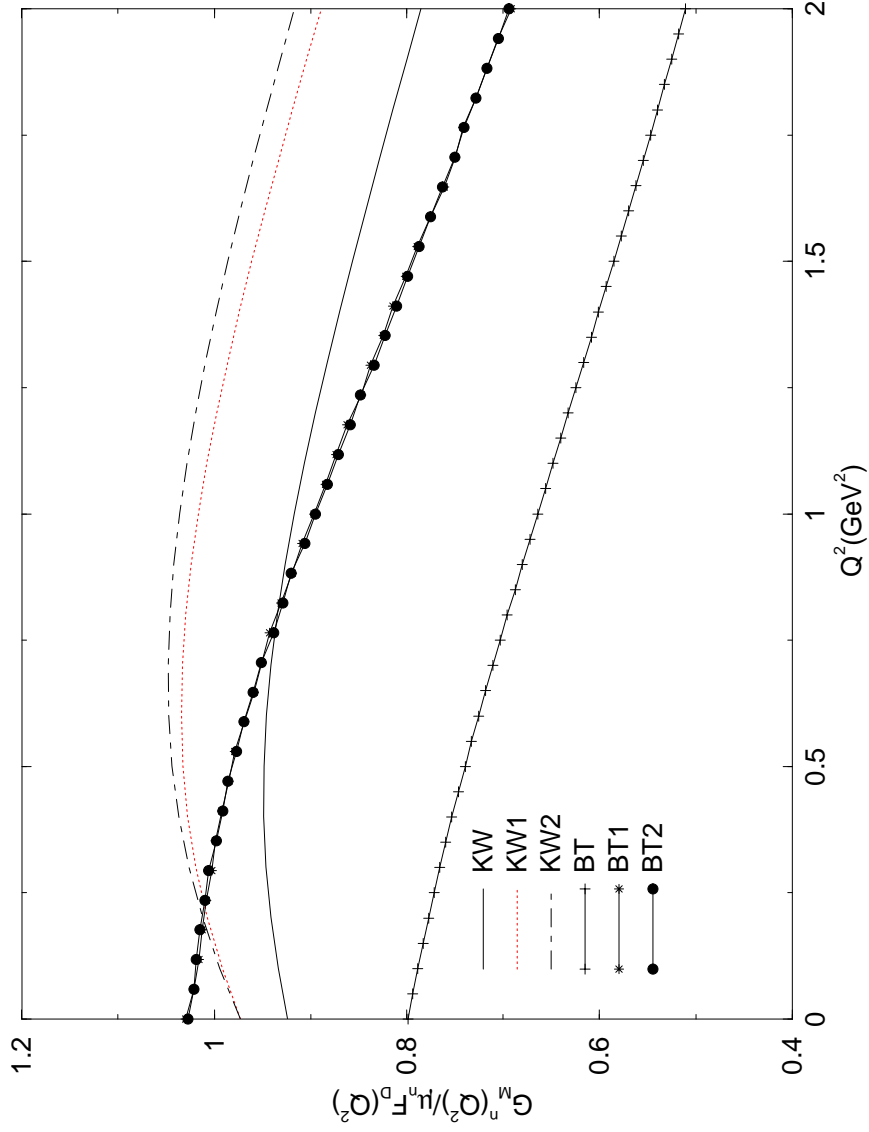,width=4in}}
\vglue 2in
\caption{Neutron magnetic form factor $G^n_M/\mu_n F_D$ relative to 
dipole shape $\mu_p F_D$ with the notation and parameters of Fig.~1.} 
\end{figure}

\begin{figure}[h]
\vglue 2in
\centerline{\psfig{figure=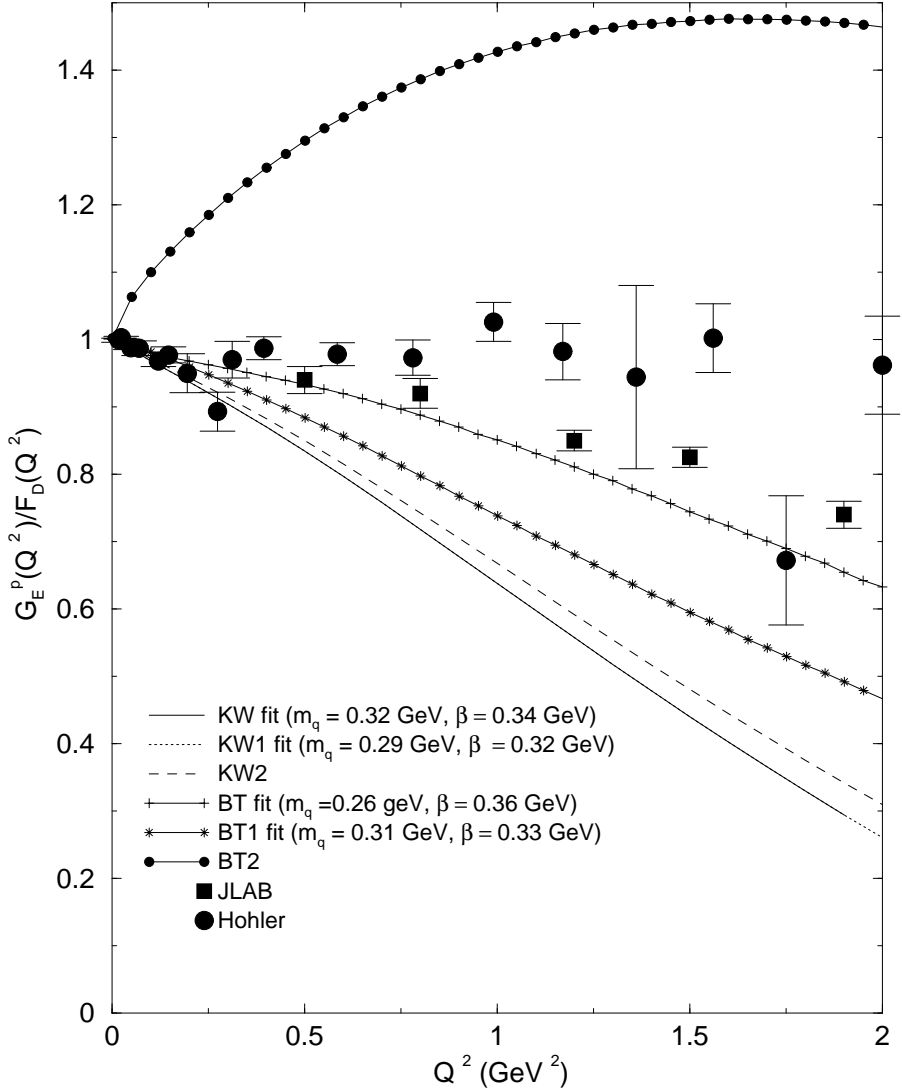,width=4in}}
\vglue 2in
\caption{Fit of $G^p_E/F_D$ for the KW-model (KW) with $m_q=0.34~\mathrm{GeV}, 
\beta=0.32~\mathrm{GeV}$ and BT-model (BT) with $m_q=0.26~\mathrm{GeV}, 
\beta=0.36~\mathrm{GeV}$ in the notation of Fig.~1. The results corrected for 
spurious form factors according to sets 1,2 are denoted by KW1, KW2 and BT1, 
BT2. The full circle data are from ref.~[21], the full squares are from~[22].} 
\end{figure}

\begin{figure}[h]
\vglue 2in
\centerline{\psfig{figure=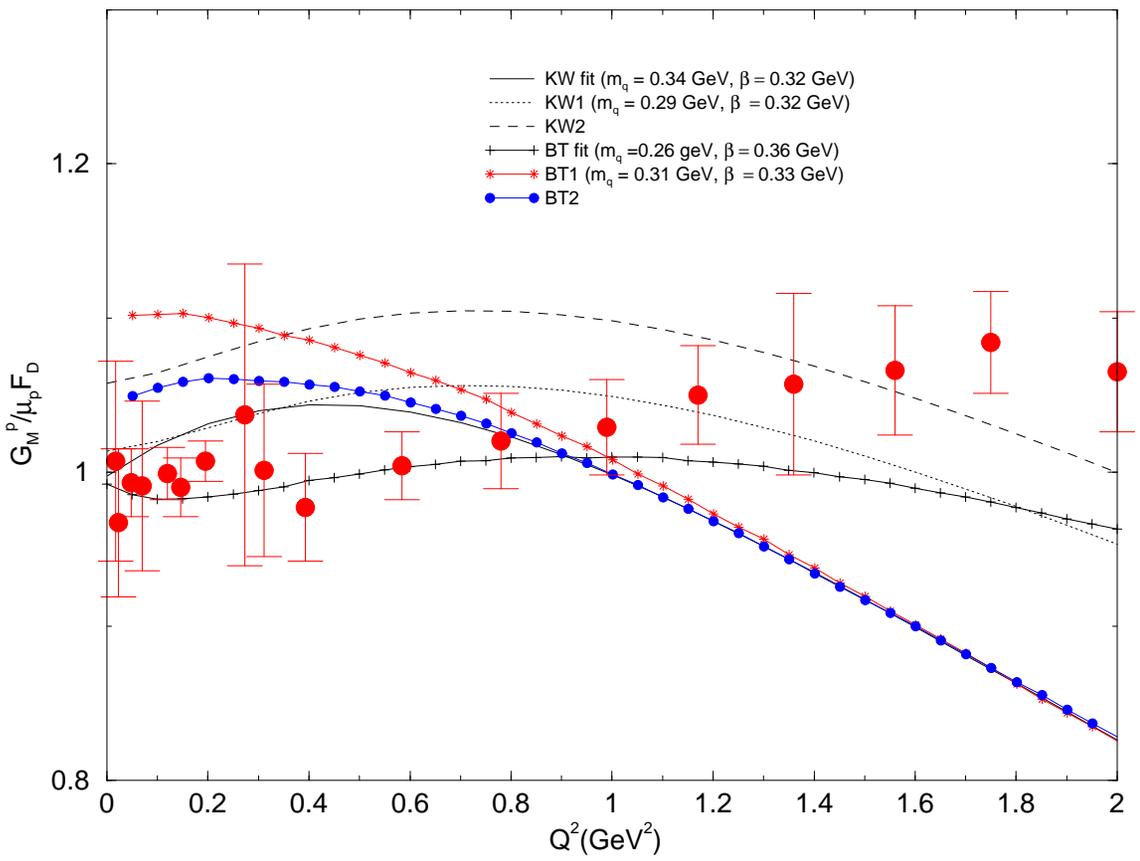,width=4in}}
\vglue 2in
\caption{ Fit of $G^p_M/F_D$ with the notation, parameters of Fig.~5. Data are 
from ref.~[21]. }
\end{figure}

\begin{figure}[h]
\vglue 2in
\centerline{\psfig{figure=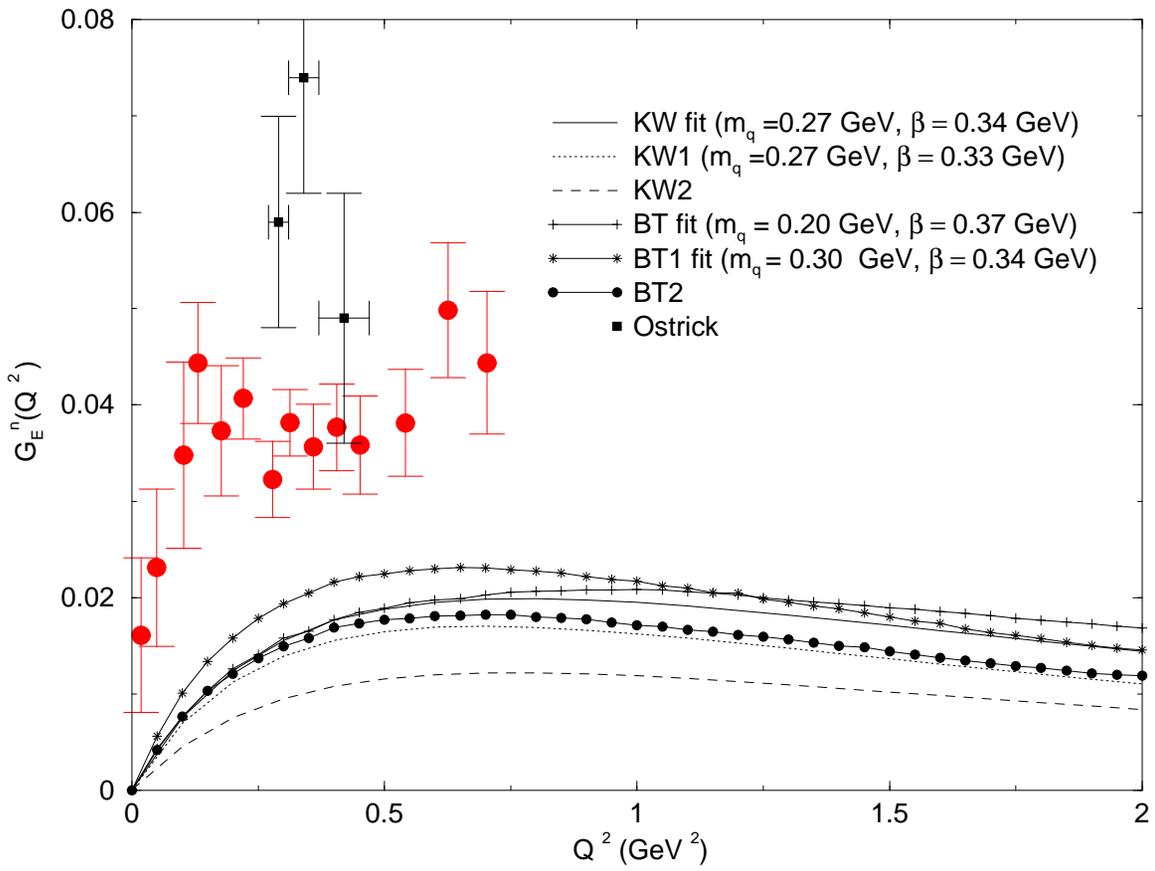,width=4in}}
\vglue 2in
\caption{ Fit of $G^n_E$ with the notation and parameters of Fig.~5. Data are 
from ref.~[23]. }
\end{figure}

\begin{figure}[h]
\vglue 2in
\centerline{\psfig{figure=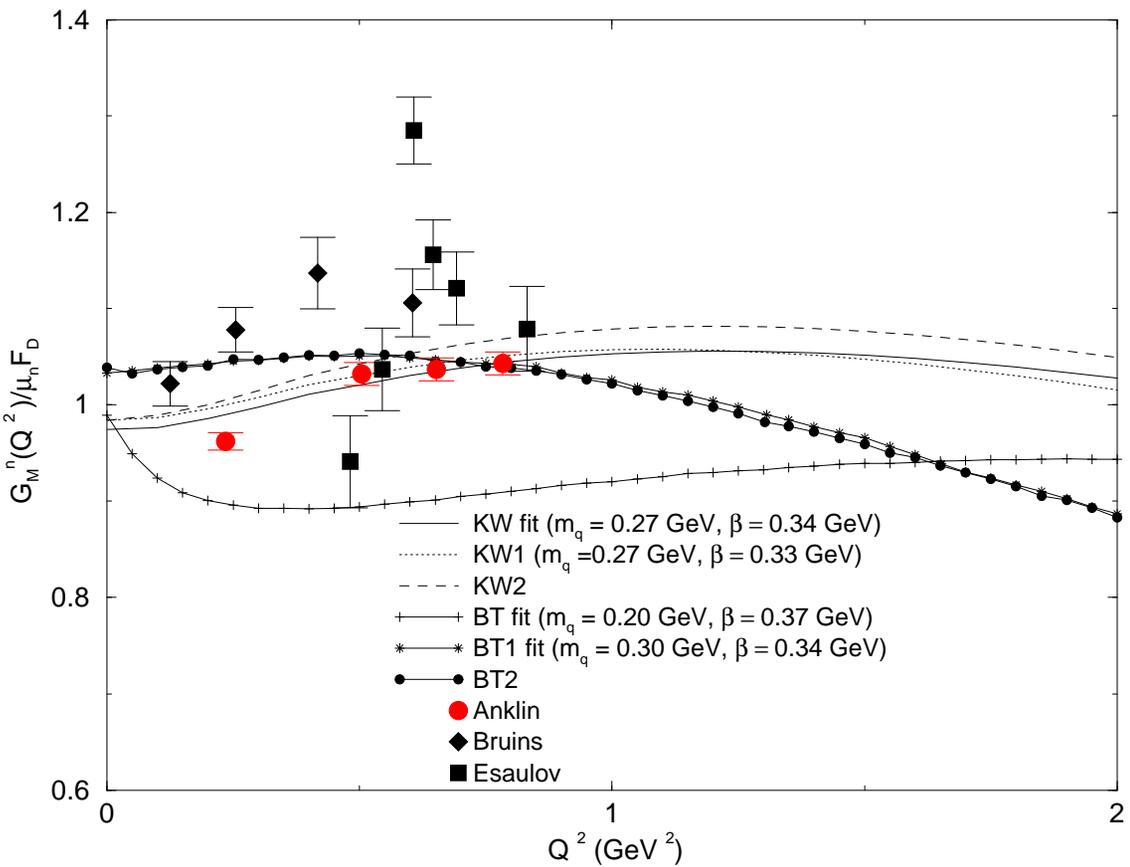,width=4in}}
\vglue 2in
\caption{Fit of $G^n_M/F_D$ with the notation and parameters of Fig.~5. The 
full circle data are from ref.~[24], the diamonds from~[25].}
\end{figure}

\begin{figure}[h]
\vglue 2in
\centerline{\psfig{figure=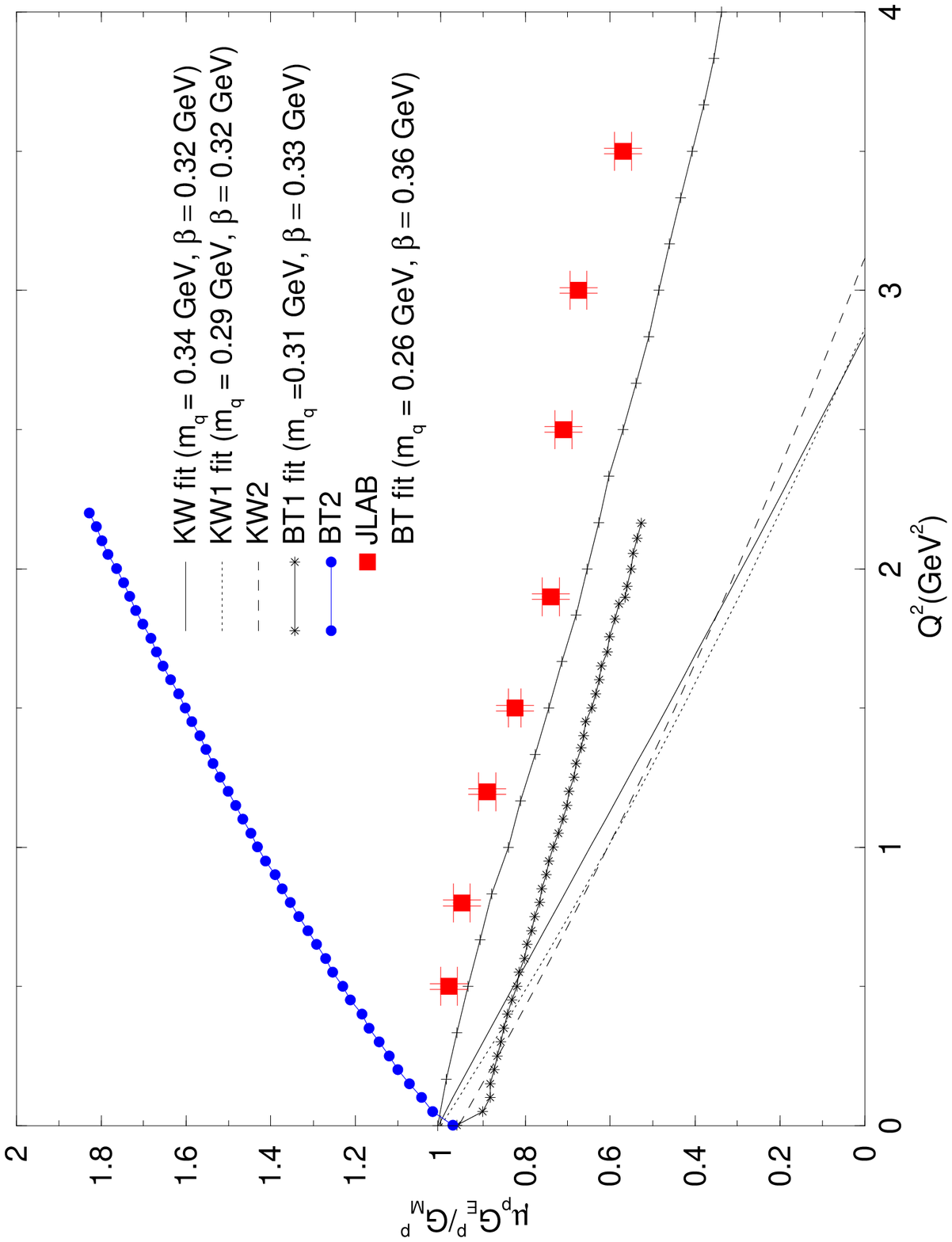,width=4in}}
\vglue 2in
\caption{Fit of $\mu_p G^p_E/G^p_M$ without spurious form factors with the 
notation and parameters of Fig.~5. The data are from JLab.~[22]. }
\end{figure}


\begin{thebibliography}{9}
\bibitem{BTel} L.~A.~Kondratyuk and M.~V.~Terent'ev, Sov. J. Nucl. 
Phys. {\bf 31}, 561 (1980); I. G. Aznauryan, A. S. Bagdasaryan and N. L. 
Ter-Isaakyan, Phys. Lett. {\bf 112B}, 393 (1982), Yad. Fiz. {\bf 36}, 1743 
(1982) [Sov. J. Nucl. Phys. {\bf 36}, 743 (1982)];  F.~Schlumpf, Phys. Rev. 
{\bf D47}, 4114 (1993); {\bf D51}, 2262 (1995); S.~J.~Brodsky and F.~Schlumpf, 
Phys. Lett. {\bf B329}, 111 (1994); F. Cardarelli and S. Simula, Phys. Lett. 
{\bf B467}, 1 (1999) and nucl-th/0006023.    

\bibitem{NPel} W. R. B. de Ara\'ujo, E. F. Suisso, T. Frederico, M. Beyer, H. 
J. Weber, Phys. Lett. {\bf B478}, 86 (2000); Z.~Dziembowski, Phys. 
Rev. {\bf D37}, 768, 778 (1988); W.~Konen and H.~J.~Weber, Phys. Rev. 
{\bf D41}, 2201(1990).   

\bibitem{PAMD} P.\ A.\ M.\ Dirac, Rev.\ Mod.\ Phys.\ {\bf 21}, 392 (1949).

\bibitem{SK} L.\ Susskind, Phys.\ Rev.\ {\bf 165}, 1535 (1968); J.\ B.\ 
Kogut and D.\ E.\ Soper, ibid.\ {\bf D1}, 2901 (1970); S.\ Weinberg, ibid.\ 
{\bf 150}, 1313 (1966). 

\bibitem{KF} V.~A.~Karmanov and A.~V.~Smirnov, Nucl. Phys. {\bf A575}, 520 
(1994); V.~A.~Karmanov, ZhETF {\bf 71}, 399 (1976)  
[Sov. Phys. JETP {\bf 44}, 210 (1976)]; ZhETF {\bf 75}, 1187 (1978) [Sov. Phys. 
JETP {\bf 48}, 598 (1978)]; M.~G.~Fuda, Ann. Phys. {\bf 197}, 265 (1990); 
{\bf 231}, 1 (1994); Phys. Rev. {\bf D41}, 534 (1990); {\bf D42}, 2898 (1990); 
{\bf D44}, 1880 (1991); M.~G.~Fuda and Y.~Zhang, Phys. Rev. {\bf C51}, 23 (1995).
 
\bibitem{FL} F.\ M.\ Lev, Nucl.\ Phys.\ {\bf A567}, 797 (1994).
\bibitem{BT} B. Bakamjian and L. H. Thomas, Phys. Rev. {\bf 92}, 1300 (1953). 
\bibitem{JF} J. Franklin, Phys. Rev. {\bf 172}, 1807 (1968). 
\bibitem{BTin} S.~Capstick and B.~D.~Keister, Phys. Rev. {\bf D51}, 3598 (1995);
F. Cardarelli {\it et al.}, Phys. Lett. {\bf B371}, 7 (1996), Nucl. Phys. 
{\bf A623}, 361 (1997).  
\bibitem{NPin} J.~Bienkowska, Z.~Dziembowski and H.~J.~Weber, Phys. Rev. Lett. 
{\bf 59}, 624, 1790 (1987); W.~Konen and H.~J.~Weber, Phys. Rev. {\bf D41}, 
2201 (1990); H.~J.~Weber, Phys. Rev. {\bf C41}, 2783 (1990); H.~J.~Weber, Ann. 
Phys. (N.Y.) {\bf 207}, 417 (1991).
\bibitem{M} H.~J.~Melosh, Phys. Rev. {\bf D9}, 1095 (1974).
\bibitem{FBW} E. F. Suisso, W. R. B. de Ara\'ujo, T. Frederico, M. Beyer and 
H. J. Weber, nucl-th/0007055. 
\bibitem{BKW} M. Beyer, C. Kuhrts and H. J. Weber, Ann. Phys.(NY) {\bf 269}, 
129 (1998). 
\bibitem{BHL} S.~J.~Brodsky, T.~Huang and G.~P.~Lepage, in {\it Quarks and 
Nuclear Forces}, eds. D. Fries and B. Zeitnitz, Springer Tracts in Modern 
Physics, Vol.100, (Springer, New York, 1982). 
\bibitem{HJW} H.~J.~Weber, Ann. Phys.(N. Y.){\bf 177}, 38 (1987). 
\bibitem{BD} We use the units $c=1=\hbar$ and $\gamma$-matrix conventions of 
J.~D.~Bjorken and S.~D.~Drell, {\it Relativistic Quantum Mechanics}, 
(McGraw-Hill, New York, 1964). 
\bibitem{KM} V. A. Karmanov and J.-F. Mathiot, Nucl. Phys. {\bf A602}, 388 
(1996).
\bibitem{LS} H.~Leutwyler and J.~Stern, Ann. Phys. (N.Y.) {\bf 112}, 94 (1978).
\bibitem{XW} Our earlier incorrect nucleon results, H. J. Weber and X. Xu, 
Nucl. Phys. {\bf A600}, 461 (1996) followed from V. A. Karmanov and A. V. 
Smirnov, Nucl. Phys. {\bf A546}, 691 (1992), {\bf 575}, 520 (1994).    
\bibitem{PL} G. P. LePage, as described in W. H. Press {\it et al.}, Numerical 
Recipes, 2nd ed., Cambridge, 1996.
\bibitem{HOE} G. H\"ohler {\it et al.}, Nucl. Phys. {\bf B114}, 505 (1976).   
\bibitem{Jlab} M. K. Jones {\it et al.}, Phys. Rev. Lett. {\bf 84}, 1398 (2000).
\bibitem{PLA} S. Platchkov {\it et al.}, Nucl. Phys.{\bf A510}, 740 (1990).
\bibitem{AN} H. Anklin {\it et al.}, Phys. Lett. {\bf B428}, 248 (1998).
\bibitem{BRU} E. E. W. Bruins {\it et al.}, Phys. Rev. Lett. {\bf 75}, 21 (1995).
\bibitem{BAR} W. Bartel {\it et al.}, Phys. Lett. {\bf B30}, 285 (1969). 

\end{thebibliography}
\end{document}